\documentstyle[epsfig,rotate]{elsart}
\begin{document}
\begin{frontmatter}
\title{Simulation of reproductive risk and emergence of female 
reproductive cessation}

\author{\bf A.O. Sousa\thanksref{email1}} 
 
\address{\it Institute for Theoretical Physics, Cologne University, 
D-50937 K\" oln, Germany}
\thanks[email1]{e-mail: sousa@thp.uni-koeln.de}

\date{\today}
\maketitle

\begin{abstract}
Using a simple computer model for evolution, we show that in a sexual 
population subject only to age-increasing reproductive risk, a
cessation of female reproduction emerges.
\end{abstract}


\end{frontmatter}
\section{Introduction}
The significance of post-reproductive survival is one of most intriguing 
evolutionary unanswered questions related to reproduction and aging. A very 
important example of that is the human menopause. Despite of several 
theories proposed, the early female reproductive senescence, which has
been also observed, although generally less clear-cut, in some other
species, for instance, rats, macaques, lions, pilot whales,
chimpanzees, elephants \cite{species} remains one of the great puzzles
of evolution theory.

In the specific case of humans, whether this postreproductive life 
is an adaptive consequence of natural selection, or is a non-adaptive 
artifact of the rapid increase in longevity over the past few centuries 
\cite{nesse}, many theories why the menopause evolved are based on 
unusual circumstances affecting the human life history (for review, see 
Ref.\cite{kirkwood1} and references therein). It has been 
suggested that menopause could act evolutionarily to protect the ageing 
woman from the hazards of childbirth. That is, as humans evolved and 
became able to reach greater ages, there came 
a point when survival during childbirth began to decline as a function 
of further ageing and increased frailty \cite{williams}. 'Premature' 
death of the mother would also put at risk any existing children and 
their potential for reproduction \cite{rochat}. At the age when the risk 
of child-bearing outweighs the benefit of producing progeny, natural 
selection would favour women who became infertile; thus, the evolutionary 
pressure for menopause. A different theory is that the human menopause arises 
from the prolonged infant dependency on the mother, coupled with the 
risks of late pregnancy and child-bearing, due to the large neonatal 
brain size. An alternative theory is that menopause enhances fitness 
by producing post-reproductive grandmothers who can assist their adult 
daughters. Recently a theoretical model incorporating both of these 
theories has been sucessful in providing a possible evolutionary 
explanation of menopause \cite{kirkwood2}.

Investigations of evolutionary problems by physicists have in fact 
boomed in the last few years. In particular, they have pioneered the 
use of techniques derived from the availability of powerful low-cost 
computers to fulfill the lack of and complement experimentation 
\cite{landau}. Since computer simulations of natural systems 
can provide much insight into their fundamental mechanisms, they can 
be used to test theoretical ideas that could be otherwise viewed 
as too vague to deserve the status of scientific knowledge \cite{book}.
Among the many computer models introduced to describe the evolution of 
populations \cite{eigen}, the Penna bit-string model \cite{penna}, 
the Redfield model \cite{redfield} and Stauffer model \cite{dresden,radomski} 
have been stood out for predicting many phenomena in population 
dynamics. 

The sexual version of the Penna model for mutation accumulation theory 
was first introduced by
Bernardes \cite{bernardes}, followed by Stauffer et
al. \cite{stauffer} who adopted a slightly different strategy. We are
going to describe the second one. The genome of each (diploid)
organism is represented by two computer words. In each word, a bit set
to $1$ at a position (``{\it locus}'') corresponds to a deleterious
mutation - a ``perfect'' strand would be composed solely of zeros. The
effect of this mutation may be felt by the individual at all ages
equal to or above the numerical order of that locus in the world. As
an example, a bit set to $1$ at the second position of one of the
bit-strings means that a harmful effect may become present in the life
history of the organism to which it corresponds after it has lived for
two time periods (``years''). In order to count the accumulated number
of mutations and compare it with a specified threshold $T$, it is
necessary to distinguish between recessive and dominant mutations. A
mutation is counted if two bits set to $1$ at the same position in
both strings or if it appears in only one of the bit-strings but at a
dominant position. The dominant positions are randomly chosen at the
beginning of the simulation and are the same for all the
individuals. Reproduction is modeled by the introduction of new
genomes in the population. Each female and male becomes able to reproduce after
having reached a minimum age, after which a female randomly chooses an
able male to breed and it generates a fixed number of offspring at the
completion of each period of life. To construct one offspring genome so, each
string of the mother genome is cut at a randomly selected position,
the same for both strings, and the left part of one is combined with
the right part of the other, thus generating two new combinations of
the original genes. Finally, $m_f$ deleterious mutations are randomly
introduced, and then the selection of one of these completes the
formation of the haploid gamete from the mother. The same process
occurs with the male's genome, producing the male gamete with $m_m$
deleterious mutations. These two resulting bit-string form the
offspring genome. The gender of the newborn is then randomly selected,
with equal probability for each sex.
 
The Redfield's model is an elegant model requiring much less computer time 
than the Penna model, but having no age structure. It is not a
population dynamics model following the lifetime of each individual,
but only simulates their probabilities to survive up to
reproduction. In Redfield's program the population is characterized by
a distribution of mutations $P(m)$, $m=0,1,...$, giving the probability
that an individual has $m$ genetic diseases in the genome. Darwinian
selection of the fittest then transforms this $P(m)$ into a survivor
distribution $L(m) \propto (1-s){^m}P(m)$ giving the probability that
a survivor has $m$ deleterious mutations in the genome, where $s=0.1$
is a selection coefficient. In the sexual case $n$ new hereditary
mutations happen according to the Poisson distribution $\mu{^n}
\exp(-\mu)/n!$, where $\mu \simeq 1$ corresponds to the mutation rate
per genome per generation. In the sexual case, after selection
transformed the progeny $P(m)$ distribution into the survivor
distribution $L(m)$, mutations of a rate $\mu$ produce the female
distribution $F(m)$ and those of a rate $\alpha\mu$ give the male
distribution $M(m)$. Then male gametes (sperm cells) are produced
containing half of the male genome; thus, their mutation number $m_m$
is roughly half of the number of the number of mutations in the
father's genome. Analogously, the number $m_f$ of mutations in the
female gametes (egg cells) is roughly half of the mother's number of
mutations. The fusion of two gametes adds these two numbers,
$m=m{_m}+m{_f}$, to produce the mutations in the progeny distribution
$P(m)$. The cycle is repeated until changes in each distribution
become negligible. In fact, the mutations in each gamete are not
exactly half of the parent but follow a binomial distribution, which
simulates the random selection of the transmitted half of the genome
via the process of meiosis, crossover, and mitosis. 

Of particular interest here is the Stauffer model, which although 
being a severe abstraction of biology, is particularly simple and had 
early successes in reproducing observed features of real populations, 
such as the Gompertz law \cite{makowiec}, the catastrophic senescence 
of Pacific Salmon \cite{ortmanns}, the vanishing of cod in the northwest 
Atlantic through over-fishing \cite{radomski}, and the social needs of humans 
for a minimum population size \cite{radomski}.

The present work can represent a step in the direction of 
understanding more the evolutionary origin of menopause. In this way, 
the simulations reported here test recent assumptions proposed in a 
work using the Penna model, in which menopause was found to self-organize 
from a combination of two effects: a reproductive risk of the mother 
that increases with the number of active mutations and thus with age; and 
child care, in the sense that young children die if their mother 
dies \cite{menop}. 

\section{Model}
In the sexual version of the Stauffer model \cite{sousa}, each individual of 
the population, which consists of males and females, is genetically 
represented by two integers, the minimum reproduction age $a_m$ and the 
genetic death age $a_d$. 

Individuals die with certainty if their age reaches $a_d$, and they die 
earlier with probability $V = N(t)/N_{max}$ at every iteration, 
where $N(t)$ is the actual population size and $N_{max}$ a so-called 
carrying capacity of the environment taking into account the limitations 
of food and space (Verhulst factor). For computer simulations this 
logistic factor has the benefit of limiting the size of the population to 
be dealt with. At every time step and for each individual a random number 
between zero and $1$ is generated and compared with $V$: if this number 
is smaller than $V$ the individual dies, independently of its age and
genetic death age $a_d$. 

If the female succeeds in surviving until the minimum reproduction age 
$a_m$, it generates, with probability $p_b$, $b$ offspring every iteration 
until death. The female randomly chooses a male to mate, with age 
also greater or equal to $a_m$. The offspring randomly inherits the 
$a_m$ and $a_d$ values from one of the parents, independently, apart 
from random changes (mutations) by $\pm 1$. The sex of the baby is 
randomly chosen, each one with probability $50\%$. This procedure is 
repeated for each of the $b$ offspring.

The probability $p_b(i)$ for individual $i$ to get $b$ children takes into 
account the well known (see e.g. \cite{tradeoff}) trade-off between 
fecundity and longevity and is the lower the larger the difference 
$a_d-a_m$ is: $$ p_b(i) \propto 1/(a_d(i)-a_m(i))$$
where we require $0 \le a_m < a_d \le a_{max}$.
We start our simulations with $a_m = 1$, $a_d = 16$ (or $a_d = 16$) and 
after thousand time steps the model shows an emergence of senescence and a 
typical age of death \cite{dresden}.

\begin{figure}[!h]
\epsfysize=10 cm {\epsfbox{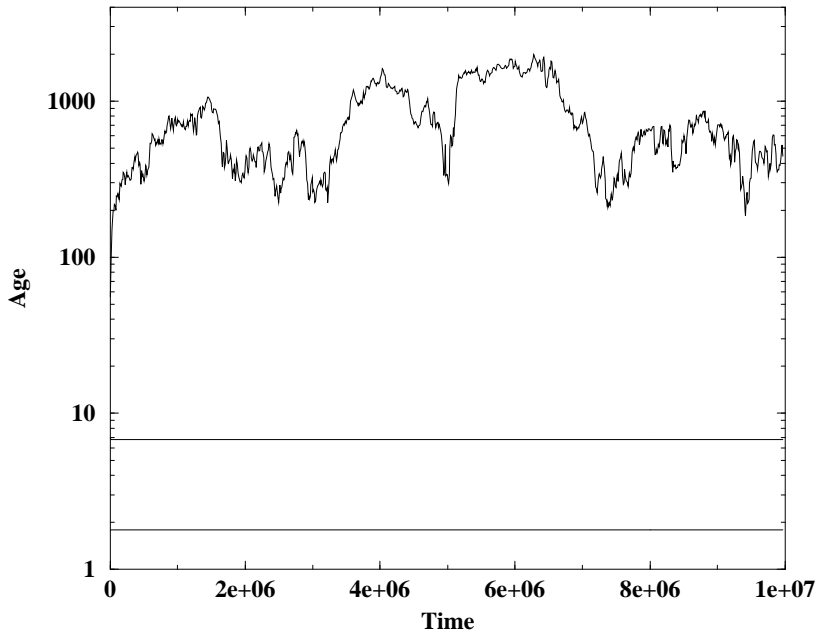}}
\caption{Time evolution of the cessation of reproduction (menopause) 
age $a_c$ (upper curve), the genetic death age $a_d$ (middle curve) and the 
minimum reproduction age $a_m$ (lower curve) for female population without 
reproductive risk. Logarithmic scales are used for the y axis (age).}
\end{figure}

In order to simulate the evolution of populations with females subjected 
only to reproductive risk $R$, we now introduce the following ingredients 
in the model:
\begin{itemize}
  \item Reproductive Risk: At the moment of giving birth, we calculate 
  the reproductive risk of a female to die. This is done through the 
  expression: 
  $$Risk = \min\left[1,\frac{\beta*(a(i)-a_m(i))}{a_d(i)-a_m(i)}\right]$$ 
  where $a(i)$ is the actual female age and $\beta$ is a predefined
  proportionality factor, which can reduce or increase the whole risk
  function. A random number between zero $r \in [0,1]$ is drawn, and
  then the females dies if and only if $r<Risk$. 

  \item Age of cessation of female reproduction $a_c$, introduced here 
  as a new variable. In the usual model all individuals can reproduce 
  at every period of life until death. Now, for females, we define a 
  maximum age of reproduction $a_c$, which is equal to $a_d$ at the 
  beginning of the simulation. It means that - {\it only at the beginning 
  of the simulation} - females can reproduce until the end of their 
  lives (there is no menopause at the beginning). When a female with a 
  given value of $a_c$ gives birth to a daughter, the daughter's value 
  of $a_c$ is mutated randomly to $a_c \pm 1$.
\end{itemize}

\section{Results}
To obtain all curves presented here we simulated $20$ different 
populations (samples) during $10,000,000$ time steps. The parameters 
of the simulations are:
\begin{itemize}
\item Initial population = $30,000$ (half for each sex);
\item Maximum population size $N_{max} = 1,500,000$;
\item Number of offspring $b = 3$;
\end{itemize}

Simulations with less samples or shorter times sometimes gave misleading 
stability. In Figure 1, which corresponds to the standard results obtained 
using the original sexual Stauffer model without reproductive risk, we 
present the time evolution of the average value of the minimum age of 
reproduction $a_m$, the genetic death age $a_d$ and the maximum age of 
reproduction $a_c$ (top to bottom, respectively). From
this figure we can observe that in order to maximize the probability
$p_b$, which happens when the difference $(a_d-a_m)$ is minimized,
$a_d$ ($\overline{a_d}\simeq6$) stays a value not much bigger than $a_m$
($\overline{a_m}\simeq2$). If the death age $a_d$ would evolve to
a very large value when compared to the minimum age of reproduction $a_m$, 
then the probability to reproduce $p_b$ would be very small, because
$p_b$ decreases for large $(a_d-a_m)$. As well, we also can see in
this figure that the menopause age becomes bigger than the genetic
death age, since $a_d$ can not be greater than $a_{max}$ and there
isn't any restriction in the maximum value of $a_c$, in such a way
that the maximum age of reproduction $a_c$ is allowed to reach values
bigger than the death age $a_d$. This means only that the females can
reproduce until the end of their lifes, i.e., the reproductive
cessation in females can not be observed in this old version without 
reproductive risk.

\begin{figure}[!h]
\epsfysize=10cm {\epsfbox{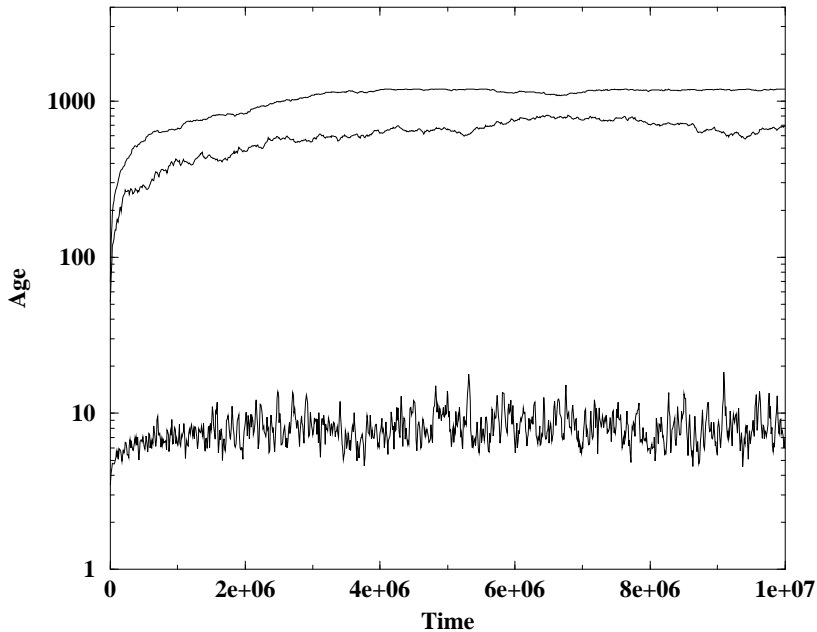}}
\caption{Time evolution of the genetic death age $a_d$ (upper curve), the 
cessation of reproduction (menopause) age (middle curve) and the minimum 
reproduction age (lower curve) for female population with reproductive risk
using $\beta=1.75$ and $a_{max}=1200$. Logarithmic scales are user for 
the y axis (age).}
\end{figure}

On the other hand, when the reproductive risk is introduced, as can be seen
in Figure 2 (with $a_{max}=1200$), the menopause age $a_c$ stays below 
the death age $a_d$. From this result we can notice
that there is a compromise between the lower reproductive risk $Risk$ and the
probability to reproduce $p_b$, which is related to the value of
the difference $(a_d-a_m)$ : if $(a_d-a_m)$ reaches a small value, i.e.,
$a_d$ not much bigger than $a_m$, a high reproduction rate dominates and
a strong selection pressure caused by the high reproductive risk could
drive the population to extinction. In this way, in order to soften the
effects from the reproductive risk and to warrant the survival of the
population, $a_d$ evolves to values much bigger than $a_m$. 

This particular feature that $a_d$ plays in the model with or without
reproductive risk have been obtained from simulations when two different
maximum values $a_{max}$ - the maximum value which $a_d$ can reach in
the simulation - were used. When we consider $a_{max}=200$ (see Fig. 3) the
difference $(a_d-a_m)$ is still big enough to produce a strong
selection pressure against the females to warrant the survival of the
population. In other words, the high reproductive risk drives the
females to remain reproducing as long as they live, it means that
they will reproduce until the end of their lifes. However, if $a_d$ is
allowed to reach a very high value (see Fig. 2, where
$a_{max}=1200$), the opposite occurs and the cessation of reproduction
of the females appears, since now the females do not suffer very
strong effects from the reproductive risk due the large difference
$(a_d-a_m)$. Given that the reproductive risk also increases with the
age, it would seem to make sense for the females to cease reproducing
at a certain age $a_c$ below the death age $a_d$ (but big enough to
guarantee the survival of the population) in order to avoid completely
the effects from the reproductive risk, since when the females stop to
reproduce there will be no reproductive risk anymore.

The emergence of female menopause age obtained here (Fig. 2), using
only a reproductive risk in giving birth that increases with age, was
found with the Penna bit-string model \cite{menop} only when in
addition to some kind of reproductive risk also maternal care was
considered. Menopause was not observed in \cite{menop} for systems
with only maternal care or with only reproductive risk. Furthermore, 
only $20\%$ of the fertile female population
had post-reproductive life and it was possible only when a very high period
of maternal care was used, which represented almost $24 \%$ of the
maximum age of life in their simulations \cite{menop}. Their simulations 
had been
performed under those restrictions suggested $40$ years ago by
Williams that menopause ``may have arisen as a reproductive adaptation
to a life-cycle characterized by senescence, unusual hazards in
pregnancy and childbirth, and a long period of juvenile dependence''
\cite{williams}. While the usual justification for the
introduction of maternal care on the menopause restricts it to
the human being, our simulations also to extend to other species, we
don't take into account maternal care in our simulations, and nevertheless 
found menopause or analogs. The human being seems to be the only species 
to which kin assist in care and provisioning of young, since orphans 
regularly can survive even in adverse environments \cite{hawkes}, our 
simulations suggest that this special human feature is not necessary for 
the emergence of menopause or analogs.

\begin{figure}[!ht]
\epsfysize=10 cm {\epsfbox{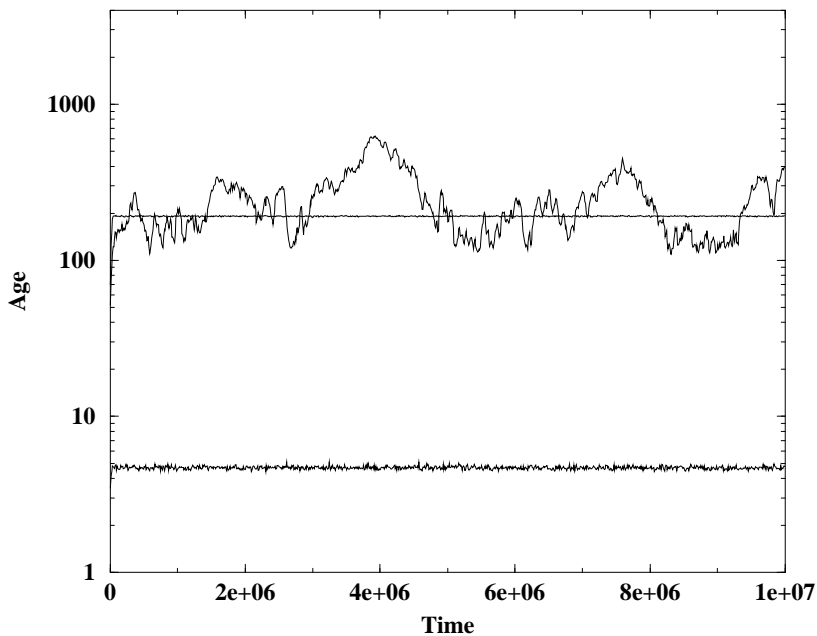}}
\caption{Time evolution of the genetic death age (flat upper curve), the 
cessation of reproduction age (fluctuating curve) and the minimum 
reproduction age (lower curve) using $\beta=1.75$ and $a_{max}=200$. 
Logarithmic scales are user for the y axis (age).}
\end{figure}

Similar results have been also observed when using different functions
for the reproductive risk, however in all cases simulated the
emergence of menopause and its analogs depends on the suitable
choice of parameters within the same model. Many simulations were
performed in order to avoid to impose a maximum value for the death age, 
$a_{max}$, but without success: without this condition it was
observed that the death age $a_d$ evolves to a fixed value below the
cessation age of reproduction $a_c$, which seems to drift towards
infinity.  

In addition, specifically concerning the constant $\beta$ which can
reduce or increase the whole risk function, we didn't manipulate it 
adequately in order to reproduce real problems related to the risks 
of childbearing the case of humans only. Instead, we had focused on 
the minimum value of $\beta$ for which would be possible
to notice the emergence of menopause using the simple Stauffer
model. Different simulations were made, as it can be seen in Table 1, 
that show stationary values of $a_m$, $a_d$ and $a_c$ averaged over a 
prespecified number of time steps. 

\begin{center}
\begin{tabular}{|c|c|c|c|c|} \hline
$\beta$ & $\overline{a_m(i)}$ & $\overline{a_d(i)}$ &
  $a_c(i)$ \\ \hline\hline
0.00  &  1.79  &  6.80      &  $\leadsto \infty$  \\ \hline
0.25  &  1.82  &  7.03      &  $\leadsto \infty$  \\ \hline
0.50  &  1.85  &  7.33      &  $\leadsto \infty$  \\ \hline
0.75  &  1.89  &  7.75      &  $\leadsto \infty$  \\ \hline
1.00  &  1.95  &  8.34      &  $\leadsto \infty$  \\ \hline
1.25  &  2.02  &  9.43      &  $\leadsto \infty$  \\ \hline
1.50  &  2.15  &  11.88     &  $\leadsto \infty$  \\ \hline
1.75  &  8.50  &  1182.81   &  633.69  \\ \hline 
2.00  &  11.18  &  1181.78   &  701.96  \\ \hline \hline
1.75  &  4.67  &  191.56    &  235.23  \\ \hline
\end{tabular}\\ [1.7ex]
\end{center}
\emph{\footnotesize  Table I.} {\footnotesize $a_m$,
  $a_d$ and $a_c$ averaged over the last time steps for different
  values of $\beta$ and using $a_{max}=1200$. In the last row, $a_{max}=200$.}

\section{Conclusions}
This paper reports on a simulation of populations under reproductive 
risk, in the framework of a simple model for biological aging. We show 
that a sexual population presents cessation of female reproduction as a 
functional characteristic of the system evolving under only reproductive 
risk. Thus simple explanations for menopause might be too simple since its 
existence in these simulations depends on the numerical values of model 
parameters.

\noindent{\bf Acknowledgments}

I would like to thank Suzana Moss and Dietrich Stauffer for discussions 
and a critical reading of the manuscript, and DAAD for financial support.

\end{document}